\documentclass[a4paper,12pt]{article} 

\usepackage{geometry,epsfig,subfigure}
\usepackage{amsmath}
\usepackage{amssymb}
\usepackage{graphics,epsfig,mydef}

\usepackage[square,sort&compress,comma,numbers]{natbib}

\usepackage[normalem]{ulem}

\usepackage[usenames]{color}
\usepackage{fancyhdr,url}   
\usepackage{setspace}  
\usepackage{multirow}   
\usepackage{rotating}    
\usepackage{colortbl}     
\usepackage{relsize} 
\usepackage{booktabs}
\usepackage{cancel}
\usepackage{enumerate}
\numberwithin{equation}{section}
\usepackage{soul} 
\usepackage{xcolor} 
\usepackage{pdflscape}
\usepackage{hyperref}
\usepackage[capitalise]{cleveref}
\usepackage{lineno}
\usepackage{graphicx}
\usepackage{etoolbox}
\usepackage{epstopdf}

\definecolor{darkblue}{rgb}{0,0,0.8}
\definecolor{darkgreen}{rgb}{0,0.5,0}

\long\def\symbolfootnote[#1]#2{\begingroup \def\thefootnote{\fnsymbol{footnote}}\footnote[#1]{#2} \endgroup} 

\renewcommand{\sin}[1]{ \text{sin}\hspace{0.0cm}\left( {#1} \right) }

\usepackage{amsthm}

\newcommand{\HRule}{\rule{0.9\linewidth}{0.2mm}}

\usepackage{nomencl}
\makenomenclature

\usepackage{etoolbox}
\renewcommand\nomgroup[1]{%
  \item[
  \ifstrequal{#1}{A}{\textit{Symbols}}{%
  \ifstrequal{#1}{B}{\textit{Greek symbols}}{%
  \ifstrequal{#1}{C}{\textit{Subscripts}}{}}}%
]}

\begin{document}
\renewcommand*{\thepage}{\arabic{page}}

\setstretch{1.3}

\begin{center}
\large
\textbf{Interplay of transport mechanisms \\during the evaporation of a pinned sessile water droplet\\}

\normalsize
\vspace{0.2cm}
Osman Akdag$^{a}$, Yigit Akkus$^{a}$, Barbaros \c{C}etin$^{b}$, Zafer Dursunkaya$^{c}\symbolfootnote[1]{e-mail: \texttt{refaz@metu.edu.tr}}\!$\\
\smaller
\vspace{0.2cm}
$^a$ASELSAN Inc., 06200 Yenimahalle, Ankara, Turkey\\
$^b$Mechanical Engineering Department, \.I.D. Bilkent University, 06800  \c Cankaya, Ankara, Turkey\\
$^c$Department of Mechanical Engineering, Middle East Technical University, 06800 \c Cankaya, Ankara, Turkey\\
\vspace{0.2cm}
\end{center}

\begin{center} \noindent \HRule \\ \end{center}
\vspace{-0.6cm}
\begin{abstract}

\noindent 

Droplet evaporation has been intensively investigated in past decades owing to its emerging applications in diverse fields of science and technology. Yet the role transport mechanisms has been the subject of a heated debate, especially the presence of Marangoni flow in water droplets. This work aims to draw a clear picture of the switching transport mechanisms inside a drying pinned sessile water droplet in both the presence and absence of thermocapillarity by developing a comprehensive model that accounts for all pertinent physics in both phases as well as interfacial phenomena at the interface. The model reveals a hitherto unexplored mixed radial and buoyant flow by shedding light on the transition from buoyancy induced Rayleigh flow to the radial flow causing coffee ring effect. Predictions of the model excellently match previous experimental results across varying substrate temperatures only in the absence of Marangoni flow. When thermocapillarity is accounted for, strong surface flows shape the liquid velocity field during most of the droplet lifetime and the model starts to overestimate evaporation rates with increasing substrate temperature.

\vspace{0.2cm}

\end{abstract}
\vspace{-0.6cm}
\begin{center} \noindent \HRule \\ \end{center}

\pagebreak

\printnomenclature

\nomenclature[A,1]{\textit{$c$}}{Molar concentration, $\rm mol \, m^{-3}$}
\nomenclature[A,1]{\textit{$c_p$}}{Specific heat capacity, $\rm J \, kg^{-1} K^{-1}$}
\nomenclature[A,1]{\textit{$D$}}{Binary diffusion coefficient, $\rm m^{2} \, s^{-1}$}
\nomenclature[A,1]{\textit{$\mathbf{g}$}}{Gravitational acceleration, $\rm m \, s^{-2}$}
\nomenclature[A,21]{\textit{$h$}}{Droplet height, $\rm m$}
\nomenclature[A,22]{\textit{$H$}}{Height of gas volume, $\rm m$}
\nomenclature[A,23]{\textit{$h_{fg}$}}{Latent heat of evaporation, $\rm J \, kg^{-1}$}
\nomenclature[A,2]{\textit{$H$}}{Height of gas volume, $\rm m$}
\nomenclature[A,2]{\textit{$k$}}{Thermal conductivity, $\rm W \, m^{-1} K^{-1}$}
\nomenclature[A,3]{\textit{$\dot{m}''_{ev}$}}{Evaporative mass flux, $\rm kg \, m^{-2} s^{-1}$}
\nomenclature[A,4]{\textit{$M$}}{Molar mass, $\rm kg \, mol^{-1}$}
\nomenclature[A,5]{\textit{$\mathbf{n}$}}{Unit vector in normal direction}
\nomenclature[A,6]{\textit{$p$}}{Pressure, $\rm Pa$}
\nomenclature[A,7]{\textit{$q^{\phi}$}}{Interfacial energy transfer rate per polar angle, $\rm W \, rad^{-1} $}
\nomenclature[A,81]{\textit{$r$}}{Radial coordinate, $\rm m$}
\nomenclature[A,82]{\textit{$R$}}{Droplet contact radius, $\rm m$}
\nomenclature[A,83]{\textit{$\mathbf{t}$}}{Unit vector in tangential direction}
\nomenclature[A,91]{\textit{$T$}}{Temperature, $\rm ^\circ C$}
\nomenclature[A,92]{\textit{$\mathbf{u}$}}{Velocity vector, $\rm m \, s^{-1}$}
\nomenclature[A,93]{\textit{$W$}}{Radius of gas volume, $\rm m$}

\nomenclature[B,1]{\textit{$\alpha$}}{Thermal diffusivity, $\rm m^{2} \, s^{-1}$}
\nomenclature[B,2]{\textit{$\gamma$}}{Surface tension, $\rm N \, m^{-1}$}
\nomenclature[B,3]{\textit{$\epsilon$}}{Emissivity}
\nomenclature[B,4]{\textit{$\theta$}}{Contact angle, $^\circ$}
\nomenclature[B,5]{$ \mu $}{Dynamic viscosity, $\rm Pa \, s$}
\nomenclature[B,6]{\textit{$\rho$}}{Density, $\unit{kg}\unit{m^{-3}}$}
\nomenclature[B,7]{\textit{$\sigma$}}{Stefan-Boltzmann constant, $\rm W \, m^{-2} K^{-4}$}
\nomenclature[B,8]{\textit{$\bar{\bar{\tau}}$}}{Stress tensor, $\rm Pa$}
\nomenclature[B,91]{$ \phi $}{Angular coordinate}
\nomenclature[B,92]{$\phi_{RH}$}{Far field relative humidity}

\nomenclature[C]{$a$}{Apex}
\nomenclature[C]{$g$}{Gas}
\nomenclature[C]{$i$}{Initial}
\nomenclature[C]{$l$}{Liquid}
\nomenclature[C]{$n$}{Normal}
\nomenclature[C]{$s$}{Droplet surface}
\nomenclature[C]{$sat$}{Saturation}
\nomenclature[C]{$surr$}{Surroundings}
\nomenclature[C]{$v$}{Vapor}
\nomenclature[C]{$w$}{Wall}
\nomenclature[C,1]{$\infty$}{Far field}

\section{Introduction}
\label{sec:intro}

Droplet evaporation plays a key role in various mechanisms encountered in nature, daily life and technological applications \cite{deegan1997,smalyukh2006,lim2009evaporation, kokalj2010}. It has been the subject of numerous studies due not only to the abundance of its applications but also its complexity \cite{erbil2012}. While the scientists aim to explore the contributing mechanisms and their underlying physics, the engineers seek the optimal solutions specific to the application where evaporating drops are utilized. Although the majority of applications include multiple evaporating droplets, which may be stationary or moving, studies usually consider a single isolated sessile droplet to cope with the inherent complexity of the problem \cite{chong2020}. Yet modeling the single sessile droplet evaporation retains its complexity due to the presence of various concurrent transport mechanisms.   

Previous studies were able to highlight certain energy transport mechanisms taking place inside the droplet as well as the gas phase surrounding it. Inside the droplet, convective energy transport contributes to conduction. Convective transport mechanisms stem from the buoyant and thermocapillary forces \cite{ruiz2002}. Since the dominance of thermocapillary forces over buoyant forces is well established \cite{lu2011,bouchenna2017}, convection mechanisms associated with thermocapillarity have been the subject of the studies \cite{ghasemi2010,askounis2017}. Thermocapillary effect refers to the pulling of the liquid from a warmer region to a colder region on the interface. This effect manifests certain flow structures depending on the type and configuration of the droplet itself. Among them, steady thermocapillary flow from the contact line to the apex, which generates a typical convection vortex, is the most common and has been demonstrated by previous theoretical studies \cite{hu2005,akkus2019iterative}. Its experimental observation was also made for various liquids \cite{zhang2002,brutin2011}. On the other hand, experimental observation of thermocapillary flow in water drops is known as challenging mostly due to the elimination of this effect by a small amount of contamination \cite{hu2005}. Yet the researchers were able to provide indirect \cite{duan2009investigation,ghasemi2010} and direct (optical observation) \cite{kita2016} findings on the presence of thermocapillary flow in water drops. Apart from steady thermocapillary flow, tangential surface tension gradient is also responsible for the formation of unsteady hydrothermal waves \cite{smith1983}. Considered as a three dimensional instability, hydrothermal waves were experimentally observed in not only non-evaporating constant thickness liquid layers \cite{riley1998,kavehpour2002}, but also evaporating droplets \cite{sefiane2008}. It was also reported that hydrothermal waves are often observed in volatile droplets \cite{zhong2017}.

Temperature gradient also exists in the normal direction due to evaporative cooling at the interface and higher temperature of the substrate if heated. A natural cause of this normal temperature gradient is convective flow instabilities associated with the density difference across the liquid. This instability mechanism creates convection cells as explained by Rayleigh \cite{rayleigh1916} and commonly referred as Rayleigh-B{\'e}nard convection owing to the early observations of B{\'e}nard \cite{benard1900} on thin liquid films. However, after four decades, Block \cite{block1956} discovered that cellular surface deformation at the free surface and associated cellular flow inside the film both observed by B{\'e}nard are triggered by the surface tension variation rather than density difference across the liquid. This instability mechanism is now named the B{\'e}nard-Marangoni convection and it replaces Rayleigh-B{\'e}nard convection in the case of thin films \cite{zhong2017,shi2017}, where Marangoni number, \mbox{$\mathrm{Ma}= -\frac{d \sigma}{dT} (T_w-T_a) h^2 /{\mu \alpha R}$}, is likely to dominate Rayleigh number, \mbox{$\mathrm{Ra}=   \rho g  \beta (T_w-T_a) h^4 / \mu \alpha R$}. A common example is the formation of B{\'e}nard-Marangoni convection cells inside pinned evaporating droplets with decreasing contact angle \cite{lu2011,bouchenna2017}.       

Outside the droplet, mass and energy are transported by the diffusion and convection. Early modeling attempts \cite{hu2002evap,ruiz2002,girard2006} neglected the convective transport associated with the buoyancy in gas phase. Later several experiments \cite{kelly2011,sobac2012pre,carle2013} revealed that diffusion-controlled evaporation models significantly underestimate the evaporation rate. Following empirical \cite{carle2016} and numerical \cite{saada2010,chen2017,akkus2020theoretical} models confirmed the underestimated evaporation rates especially in the case of elevated substrate temperatures.

Resolving the interfacial phenomena at liquid-gas interface is a must, since they dictate the evaporation rate of droplets both directly and indirectly. For instance, estimation of evaporation rate is directly linked to the Stefan flow of air, which opposes the diffusion of air towards the interface, through which air cannot penetrate due to its insolubility in liquid \cite{zhang2020}. Experimental studies \cite{kabov2017,misyura2018} as well as empirical and numerical models \cite{carle2016,akkus2020stefan} revealed that omission of Stefan flow results in a considerable underestimation of evaporation rates. Besides, surface forces such as the aforementioned thermocapillary forces, alter the internal flow, which leads to a significant variation in evaporation rates \cite{lu2011,akkus2020theoretical}. Energy and mass transfers are inherently coupled at the interface. Majority of the heat transfer is associated with the evaporating mass. However, other mechanisms, such as the conduction to the gas phase and radiative heat transfer to the surroundings, contribute to the interfacial energy transport especially in cases with large temperature differences.

As highlighted, droplet evaporation comprises complex and concurrent transport mechanisms in both phases. Beyond this, a bigger challenge is the switching of the mechanisms upon the variation of the configuration and the geometry of the problem. Consequently, it is almost impossible to build an assumption-free model, even for a single sessile droplet, enabling the capture of all relevant physics regardless of the changes in problem conditions. Therefore, certain simplifying assumptions have to be made during the modeling. A pre-eminent assumption of droplet modeling is the axial symmetry, which renders the problem a 2-D configuration, thereby reducing the computational cost significantly. However, this assumption prevents the capture of 3-D instability patterns such as oscillating/travelling hydrothermal waves. But axially symmetric models can still be considered as a powerful analysis tool, since they were able to produce multicellular flow pattern of B{\'e}nard-Marangoni convection in many studies \cite{lu2011,bouchenna2017,chen2017}. Outer boundary conditions for the gas phase (far field boundary conditions) are also subjected to certain assumptions. Common treatment is to include a very large gas domain in models; yet, if a certain experiment is aimed to be simulated, the inclusion of real physical boundaries such as the environmental chamber is desirable to eliminate the uncertainty associated with the outer boundary selection. On the other hand, in the case of drying droplets, the process is inherently transient. Geometry of the droplet changes with time in three modes in the following order: i) constant contact radius, decreasing contact angle, ii) constant contact angle, decreasing contact radius, iii) decreasing both contact angle and contact radius \cite{picknett1977}. However, transition between the stages mostly depends on the contact angle hysteresis due to surface roughness and surface chemistry alterations \cite{bourges1995}. Moreover, superhydrophobic or superhydrophilic surfaces cannot be subjected to anticipated evaporation stages due to excessive sliding or pinning characteristics of the surfaces. Despite the dynamic nature of droplet evaporation, a very useful and widely utilized assumption by the models is the quasi-steady-state approximation, which is justified by the fact that deformation speed of the droplet boundary is much less than the characteristic convection velocities in both phases \cite{bouchenna2017}.

In the present study, we aim to build a comprehensive model for the investigation of evaporation of drying sessile droplets by extending our recent work \cite{akkus2020theoretical}, where a theoretical framework was developed to model the evaporation from steadily fed droplets. The current model utilizes quasi-steady-state approximation and connects discrete droplet configurations of decreasing contact angle. The model incorporates all pertinent transport mechanisms in both phases such as buoyant and thermocapillary convections, Stefan flow, and species diffusion. Evaporation rate, evaporative cooling thereof, is estimated based on the concentration field of the vapor and flow field of the gas. A distinctive feature of the model is the utilization of temperature dependent thermophysical properties, which enables solving full compressible Navier-Stokes equations. Prior studies \cite{saada2010,pan2013,carle2016,bouchenna2017,chen2017} adopted Boussinesq approximation, yet this simplification was shown to lead to a considerable underestimation of evaporation rates \cite{akkus2020theoretical}.

The computational model developed is utilized for the simulation of the evaporation of a water droplet resting on a flat,  thermally highly conductive substrate subjected to different heating loads. The substrate surface is assumed hydrophilic, which favors  evaporation \cite{shanahan2011,nguyen2012}, and the evaporation mode is considered as one with a constant contact radius, since the transition to other modes is  a function of substrate surface properties, which may even change with time and hardly controllable. The model is validated across the experiment set of Sobac and Brutin \cite{sobac2012pre}, which was frequently utilized by the numerical studies \cite{chen2017,kumar2018,pan2020} for benchmark purposes.

The modular structure of the computational model enables the assessment of individual role of different transport mechanisms. Owing to this benefit, much disputed role of thermocapillarity instigated transport in droplets can be highlighted by simply switching on/off the Marangoni effect in the algorithm. To this effect, the role of Marangoni convection inside the droplets experimented by Sobac and Brutin \cite{sobac2012pre} is elucidated based on the droplet life in the presence and absence of thermocapillarity. One motivation of the current investigation is indeed the agreement of the predictions of previous models with the experimental results reported in \cite{sobac2012pre}, although active convective transport mechanism inside the droplet was thermocapillarity (and buoyancy) in \cite{chen2017} and sole buoyancy in \cite{kumar2018,pan2020}. 

\section{Computational Model}
\label{sec:modeling}

Recently presented iterative modeling approach \cite{akkus2020theoretical} is applied to model the drying droplets, whose contact area with the substrate is unchanging. Quasi-steady-state successive simulations are conducted for discrete droplets with reducing contact angles to mimic the shrink of a sessile droplet. Contact angle reduction step is constant and selected sufficiently small to secure its ineffectuality on the global results. The interlink between the discrete droplets is established by imposing the surface velocity associated with the change of the droplet geometry upon drying. Droplet preserves its spherical cap shape due to the sufficiently small initial size of the droplet such that capillary forces always dominate the gravitational ones.

\begin{figure}[h]
\includegraphics[scale=1.2]{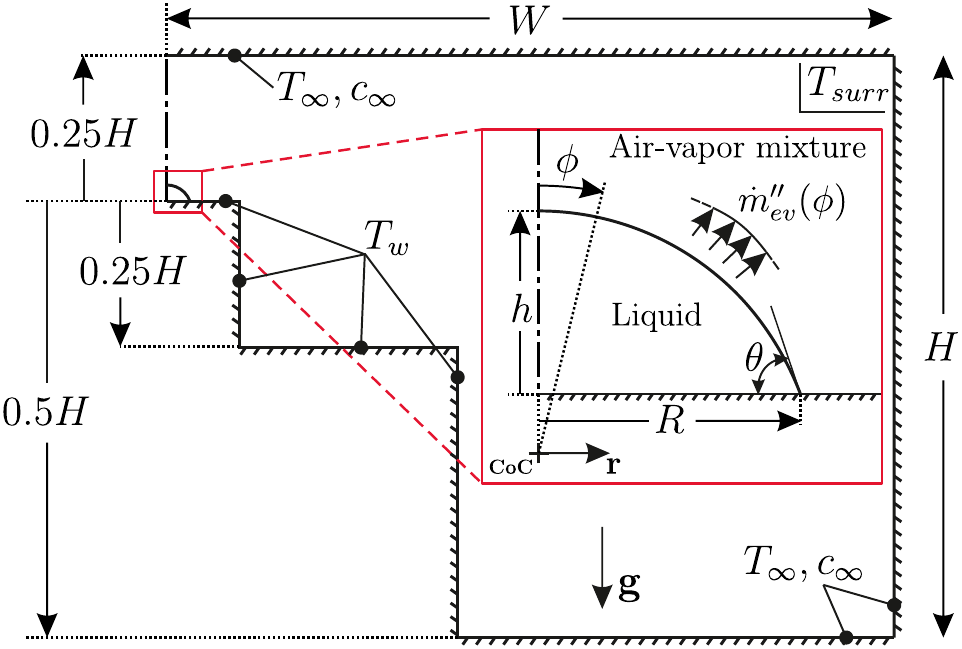}
\centering
\caption{Problem domain based on axisymmetry. Droplet is placed onto a two-stage substrate, which is assumed sufficiently thermally conductive such that walls of the substrate possess the same temperature, $T_w$. The substrate is surrounded by an environmental chamber, whose walls have a constant far field temperature, $T_{\infty}$. The gas adjacent to outer walls is assumed to have a constant far field vapor concentration, $c_{\infty}$, while, at the walls of the substrate with temperature control, gradient of vapor concentration is assumed zero. $\phi$ is the polar angle measured from the apex with respect to center of curvature (CoC) of the droplet.}
\label{fgr:domain}
\end{figure}

Problem domain is constructed in accordance with the experimental conditions in \cite{sobac2012pre} as shown in \cref{fgr:domain}. Although the original setup is surrounded by an environmental chamber in the shape of a rectangular prism, the chamber is modeled as a cylinder with the same volume due to the 2-D axisymmetric approach adopted in the present study. Height of the cylinder is  equal to that of the original environmental chamber in order to include the boundary effect associated with the upper wall of the chamber. It should be noted that the effect of physical boundaries was not included in previous modeling works \cite{chen2017,kumar2018,pan2020} focusing on the same experiment.

\subsection{Governing Equations}

Since the surface velocity associated with the shrinking of the droplet is much smaller than the convection velocities in both liquid an gas domains, quasi-steady-state approximation is applied in the modeling. Except transient terms, full compressible Navier-Stokes equations and conservation of energy equation are solved in the entire domain. In addition, in the gas domain, species conservation equation is solved for vapor transport. The computational model utilizes temperature dependent properties throughout the domain, which eliminates the need for applying any approximation to model buoyant and thermocapillary effects. Governing equations are summarized as follows:

\begin{subequations}
\begin{equation} \label{eqn:vector-cont}
\nabla\cdot(\rho\mathbf{u}) = 0
\end{equation}
\begin{equation} \label{eqn:vector-mom}
\rho(\mathbf{u}\cdot\nabla)\mathbf{u} = -\nabla p + \nabla \cdot \bar{\bar{\tau}} + \rho\mathbf{g}
\end{equation}
\begin{equation} \label{eqn:vector-en}
\rho c_p\mathbf{u}\cdot\nabla T = \nabla \cdot (k\nabla T) + \bar{\bar{\tau}}:\nabla \mathbf{u}
\end{equation}
\begin{equation} \label{eqn:vector-species}
\mathbf{u}\cdot\nabla c_v = \nabla \cdot (D\nabla c_v)
\end{equation}
\end{subequations}

\noindent where $\rho$, $c_p$, $k$, and $D$ are density, specific heat, thermal conductivity, and   
binary diffusion coefficient of the fluid, respectively; $\mathbf{u}$, $p$, and $T$  are velocity, pressure, and temperature of the fluid, respectively; $c_v$ is the molar concentration of the vapor in the gas phase; $\bar{\bar{\tau}}$ is the deviatoric stress tensor defined as \mbox{$\mu ( \partial u_{i} / \partial x_{j} + \partial u_{j} / \partial x_{i} )$}; and $\mathbf{g}$ is the gravitational acceleration. The values of the temperature dependent properties are taken from the material library of COMSOL Multi-physics software.

\subsection{Initial and Boundary Conditions}

Initial contact angle of the droplet, $\theta_i=68^{\circ}$, is the wetting angle of water droplet on the aluminium substrate coated with SiO$_{\textrm{x}}$ (\cite{sobac2012pre}). Symmetry boundary condition is applied along the center line of the axisymmetric domain. No slip boundary conditions are used on walls of the substrate and environmental chamber. Temperature distributions on the walls as well as the molar concentration related conditions at the boundaries  are specified in \cref{fgr:domain}. 

In order to determine boundary conditions at the free surface, mass, tangential force, and energy balances should be properly established at the liquid-vapor interface. Normal component of the interfacial liquid velocity ($\mathbf{u_{\ell}}$) is determined by the evaporating mass flux: \mbox{$(\mathbf{u_{\ell}}-\mathbf{u_{s}}) \cdot \mathbf{n}=\dot{m}''_{ev} / \rho$}, where $\mathbf{u_{s}}$ is the surface velocity associated with the shrinkage of the droplet. Estimation of evaporating mass flux is critical, since it plays a vital role in both mass and energy balances. Evaporating mass flux is estimated based on the additive effects of diffusive and convective interfacial transports:

\begin{equation} \label{eqn:evap_flux}
\dot{m}''_{ev}= -D  (\nabla \cdot  \mathbf{n})c_v  + (\mathbf{u_g} \cdot \mathbf{n}) c_v \, ,
\end{equation}

\noindent where $\mathbf{u_g}$ is the interfacial gas velocity; therefore, estimation of evaporation flux requires not only the solution of the concentration field of vapor in the gas domain, but also the interfacial gas velocity. To this effect, an interrelation between the vapor concentration and gas velocity should be provided at the interface. This link is actually created by the Stefan flow, which is basically the flow of air from the interface. Stefan flow originates in order to oppose the diffusion of air towards the interface, through which air cannot penetrate due to its insolubility in liquid. This phenomenon can be mathematically expressed by equating the diffusive air transport towards the interface and the convective air transport from the interface as follows:

\begin{equation} \label{eqn:stefan}
D  (\nabla \cdot  \mathbf{n})c_{air} = (\mathbf{u_g} \cdot \mathbf{n}) c_{air} \, ,
\end{equation}

\noindent where $c_{air}$ is the molar concentration of air. Then \Cref{eqn:stefan} can be used to estimate the normal component of the gas velocity, \textit{i.e.} $u_n=(D/c_{air})(\partial c_{air}/ \partial n)$, which enables the calculation of evaporating mass flux (\Cref{eqn:evap_flux}) for a given concentration field. 

Estimation of the tangential velocity, which is the same for both phases \cite{prosperetti1979,carle2016}, is obtained from the tangential stress balance at the interface. Tangential stress balance expresses the interplay of thermocapillary forces arising from the non-uniform distribution of the interfacial temperature and shear forces induced on the interface by the two phases:

\begin{equation} \label{eqn:force_tangent}
\mathbf{n} \cdot {\bar{\bar{\tau}}}_g \cdot \mathbf{n} -\mathbf{n} \cdot {\bar{\bar{\tau}}}_{\ell} \cdot \mathbf{t} = \nabla\gamma \cdot \mathbf{t} \, ,
\end{equation}

\noindent where $\gamma$ is the surface tension, and $\mathbf{t}$ is the unit vector in tangential direction. Based on the assumption of the shear force associated the gas being much smaller than that of the liquid, the effect of gas shear on the interface force balance is neglected. This assumption is validated by an \textit{a posteriori} analysis of the results. 

Energy exchange at the interface of an evaporating droplet involves several physical mechanisms. The major one, the evaporative heat transfer, is associated with the breaking of physical bonds between liquid molecules. Diffusive heat transfer, conduction, accompanies the evaporation. Yet its direction may be reversed wherever the gas temperature exceeds the liquid surface, which is possible in the case of heated substrates. Another mechanism, thermal radiation, is associated with the emission of electromagnetic waves. It always exists between the interface and surroundings. Then interfacial energy balance can be expressed as follows: 

\begin{equation} \label{eqn:energy_balance}
\mathbf{n} \cdot (-k_l \nabla T_l) = \dot{m}''_{ev}h_{fg}-\mathbf{n} \cdot (-k_g \nabla T_g) 
+\sigma \epsilon (T_s^4-T_{surr}^4) \, ,
\end{equation}

\noindent where $h_{fg}$, $\sigma$, and $\epsilon$ are latent heat of vaporization, Stefan-Boltzmann constant, and emissivity of the liquid surface, respectively. The subscripts $s$ and $\mathit{surr}$ designate the droplet surface and surroundings, respectively. Temperature of the surroundings is assumed equal to the temperature of the environmental chamber. 

Finally, boundary conditions for the liquid and gas domains are summarized in Eqs.~(\ref{eqn:liq-b1})~to~(\ref{eqn:liq-b4}) and Eqs.~(\ref{eqn:gas-b1})~to~(\ref{eqn:gas-b4}), respectively, as follows:  

\begin{subequations}
\begin{equation} \label{eqn:liq-b1}
\partial_\phi \mathbf{u} =0 \hskip 1 pt ; \hskip 3 pt  \partial_\phi T=0 \hskip 5 pt {\rm at} \hskip 5 pt \phi=0
\end{equation}
\begin{equation} \label{eqn:liq-b2}
\mathbf{u}=0 \hskip 1 pt ; \hskip 3 pt T=T_w \hskip 5 pt {\rm at} \hskip 5 pt  \phi=\theta
\end{equation}
\begin{eqnarray} \label{eqn:liq-b4}
&(\mathbf{u}-\mathbf{u_s}) \cdot \mathbf{n}=\dot{m}''_{ev}/{\rho} \hskip 1 pt , -\mathbf{n} \cdot \bar{\bar{\tau}} \cdot \mathbf{t}=\nabla\gamma \cdot \mathbf{t} \hskip 1 pt ; \nonumber \\
&\mathbf{n} \cdot (-k_l \nabla T) = \dot{m}''_{ev}h_{fg}-\mathbf{n} \cdot (-k_g \nabla T_g) 
+\sigma \epsilon (T^4-T_{surr}^4) \hskip 5 pt {\rm at} \hskip 5 pt  r=R/ \sin\theta
\end{eqnarray}
\end{subequations}

\begin{subequations}
\begin{equation} \label{eqn:gas-b1}
\partial_\phi \mathbf{u} =0 \hskip 1 pt ; \hskip 3 pt  \partial_\phi T=0 \hskip 1 pt ; \hskip 3 pt \partial_\phi c_v=0 \hskip 5 pt {\rm at} \hskip 5 pt \phi=0
\end{equation}
\begin{equation} \label{eqn:gas-b2}
\mathbf{u}=0 \hskip 1 pt ; \hskip 3 pt T=T_w \hskip 1 pt ; \hskip 3 pt \partial_\phi c_v=0  \hskip 5 pt {\rm on} \hskip 3 pt  {\rm the} \hskip 3 pt  {\rm heated} \hskip 3 pt {\rm walls}
\end{equation}
\begin{equation} \label{eqn:gas-b3}
\mathbf{u}=0 \hskip 1 pt ; \hskip 3 pt T=T_{\infty} \hskip 1 pt ; \hskip 3 pt c_v= c_{\infty}=\phi_{RH} c_{v,sat} \hskip 5 pt {\rm on} \hskip 3 pt  {\rm the} \hskip 3 pt  {\rm chamber} \hskip 3 pt {\rm walls}
\end{equation}
\begin{equation} \label{eqn:gas-b4}
\mathbf{u}=\mathbf{u_g} \hskip 1 pt ; \hskip 3 pt T=T_s  \hskip 1 pt ; \hskip 3 pt c_v=c_{v,sat}  \hskip 5 pt {\rm at} \hskip 5 pt  r=R/ \sin\theta
\end{equation}
\end{subequations}

\noindent where $\phi_{RH}$ is the far field relative humidity and $c_{v,sat}$ is the saturation concentration of vapor at the corresponding temperature. Because of the assumption of thermal equilibrium, surface temperature calculated from the solution of liquid domain is assigned to the gas domain. Distribution of the interfacial gas velocity ($\mathbf{u_g}$) is based on the normal component, which is calculated from the Stefan flow (see \cref{eqn:stefan}), and tangential component, which is estimated from the solution of the velocity field in the liquid domain.

\subsection{Solution Methodology} 
\label{sec:scheme}

Computational model presented in the current work solves the governing equations in both phases, separately. However, boundary conditions are inseparably interconnected at the droplet surface because of the concurrent interfacial phenomena affecting both phases such as the conjugate heat and mass transfer and Stefan flow. Therefore, coupling of the phases should be carefully handled. To this effect, the present model utilizes an iterative computational scheme, in which liquid and gas domains are successively solved utilizing the Finite Element Method (FEM) based solver of COMSOL Multiphysics software. The iterative scheme is implemented using the interface, Livelink for MATLAB. In FEM formulation, variable discretization is implemented by linear shape functions in both domains and for all variables. 

Computational domain is meshed by the mesher of COMSOL itself. Mesh generation initiates at the interface, which is divided to arcs of equal length, and continues towards the liquid and gas domains with a certain growth rate. Resolution of the solution, thereby the size of the mesh, is controlled by the size (length) of the arcs at the interface. The same arc length is utilized at all contact angles. Arc length independence test, which can be viewed as a mesh-independence analysis, is performed by utilizing diminishing values for the arc length. In terms of droplet life, simulations yield almost identical results for all cases with the maximum relative error of 0.5\%. 

Another parameter that may influence the global results is the reduction step of contact angle between the drying simulations. Contact angle reduction step is selected as 4$^\circ$ for all simulations. The effect of this selection is assessed by comparing the results of simulations with the reduction step of 1$^\circ$ for selected cases. In terms of droplet life, the change of the result is always less than 0.7\%.

\section{Results and Discussion}
\label{sec:results}

The proposed model is applied to simulate the evaporating water droplet experiments in \cite{sobac2012pre}. The physical configuration shown in \cref{fgr:domain} is identical to the experiments except the rectangular prism shapes of the heater block and environmental chamber. Yet, the heights are the same with those in experiments. In addition, radii are selected based on the average of the lateral dimensions. Three different substrate temperatures are considered: one corresponds to the isothermal substrate case and the others are higher than the ambient. The values of the geometric parameters and simulation conditions are summarized in \cref{table:props}. Emissivity of the water surface taken as 0.97 (\cite{robinson1972}). Vapor-air diffusion coefficient is calculated based on the temperature dependent formulation suggested in \cite{bolz1976}. All other thermophysical properties are also temperature dependent and their values are taken from the material library of COMSOL.

\begin{table}[h]
\caption{Geometrical parameters and simulation conditions}
\begin{center}
\begin{tabular}{lll}
\hline 
Droplet contact radius (mm)             & $ R $    & 1.44 \\    
Initial contact angle ($^\circ$)          & $ \theta_i $   & 68 \\ 
Radius of gas volume (mm)       & $ W $      & 500 \\ 
Height of gas volume (mm)       & $ H $      & 400 \\ 
Far field relative humidity       & $ \phi_{RH} $       & 0.475 \\ 
Far field temperature ($^{\circ}$C)  & $\rm{T_{\infty}}$   & 25.4 \\ 
Substrate temperatures ($^{\circ}$C)  & $\rm{T_w}$   & 25.4, 55.4, 65.4 \\ 
\hline 
\end{tabular} 
\end{center}
\label{table:props}
\end{table}

Transport mechanisms inside the droplet directly affect the evaporation rates, or the droplet lifetime thereof. When Marangoni flow is present, it is responsible for the majority of mass and energy transport. In its absence, buoyant flow and/or radial flow can be effective for the convective transport. On the other hand, the presence of Marangoni flow in water droplets is contentious in the literature due to water being prone to contaminants. Therefore, the current study carries out the simulations considering both scenarios: i) without the presence of Marangoni flow (w/o MA) and ii) with the presence of Marangoni flow (w/ MA). Droplet lifetime predictions of the model together with the experimental results of \cite{sobac2012pre} are presented in \cref{table:rates}. \begin{table}
\centering
\caption{Droplet lifetimes (in seconds).}
\label{table:rates}
\begin{tabular}{lccc}\\
\toprule      
$\rm{T_w}$ & 25.4\unit{$^{\circ}$C} & 55.4\unit{$^{\circ}$C} & 65.4\unit{$^{\circ}$C} \\ \midrule
Experiments in \cite{sobac2012pre}  & 1585 & 165 & 99  \\
Model w/o MA (deviation)  & 1581 ($-0.2$\%) & 169 (2.4\%) & 99 ($-$)   \\
Model w/ MA  (deviation) & 1567 ($-1.1$\%)  & 156 ($-5.5$\%) & 88 ($-11.1$\%)  \\
\bottomrule
\end{tabular}
\end{table}
Results demonstrate that the model is in excellent agreement with the experiments when Marangoni effect is omitted. Similar agreement was reported by a previous modeling attempt \cite{pan2020}, where the effect of thermocapillarity was not taken into consideration. The inclusion of thermocapillary flow, on the other hand, results in underestimated droplet lifetimes. While this deviation is prominent at elevated substrate temperatures, the gap becomes narrower with decreasing superheat. This behaviour could explain a former study \cite{chen2017}, where thermocapillary effect was accounted for in the simulations and the predictions of the model was reported to match the experimental results for isothermal substrate. Consequently, the model presented in the current study reveals that buoyant and radial flow mechanisms are sufficient for the mass and energy transport in droplets experimented in \cite{sobac2012pre}. This claim requires a close inspection of the results. In what follows, results of the simulations are further investigated to highlight the underlying physical mechanisms in both phases.

\begin{figure}[h]
\includegraphics[scale=0.7]{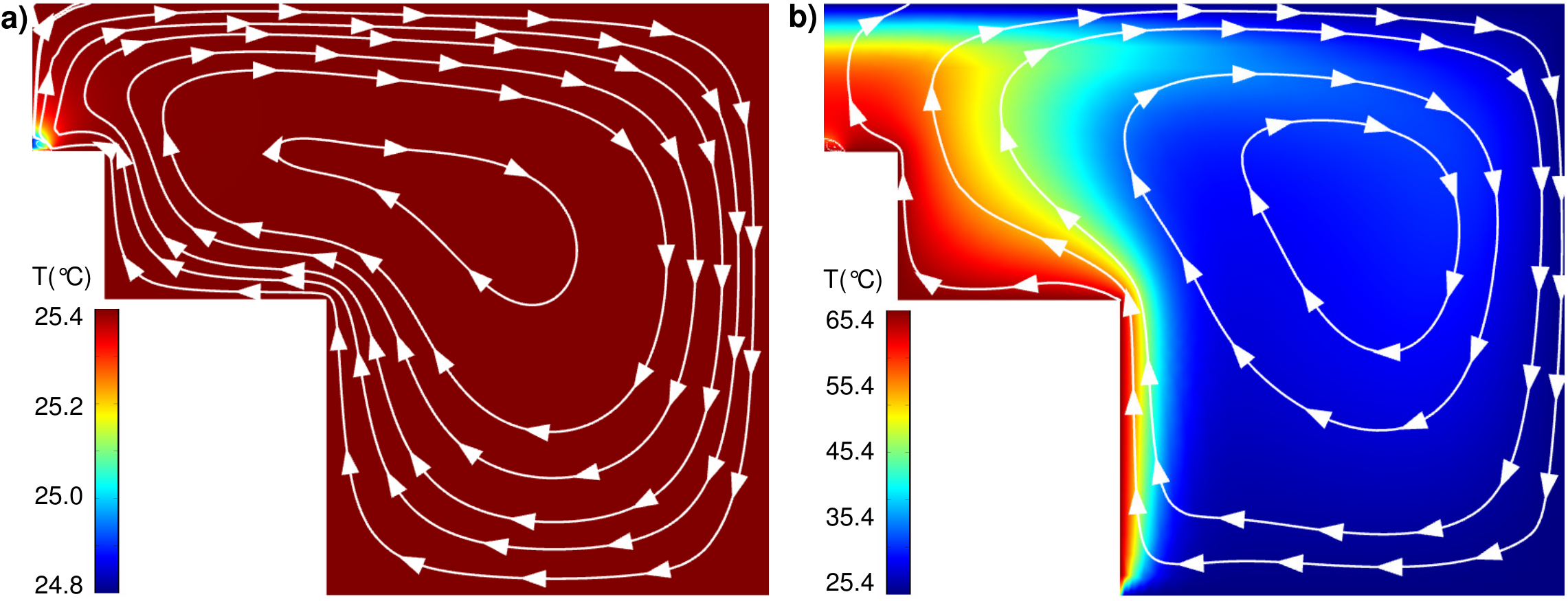}
\centering
\caption{Temperature field and streamlines in the entire computational domain for \textbf{a)} isothermal \mbox{($\rm{T_w}$=25.4\unit{$^{\circ}$C})} and  \textbf{b)} heated \mbox{($\rm{T_w}$=65.4\unit{$^{\circ}$C})} substrates.}
\label{fgr:full_T}
\end{figure}

A unique feature of the present study is the realistic modeling of the surrounding gas flow by incorporating the actual physical boundaries of the test chamber. Figure~\ref{fgr:full_T} shows the resultant temperature and velocity fields for isothermal \mbox{($\rm{T_w}$=25.4\unit{$^{\circ}$C})} and heated \mbox{($\rm{T_w}$=65.4\unit{$^{\circ}$C})} substrates. In the isothermal case, temperature distribution is homogeneous except inside and near the droplet, where evaporative cooling decreases the temperature values. In the case of heated substrate, a thermal boundary layer forms on the walls on hot walls. Away from the walls, gas temperature is close to its ambient value. Despite the different temperature fields, gas flow is quite similar in both cases. The gas moves in the clockwise direction and forms a large single convection cell and this pattern is not affected by the presence of thermocapillarity. However, the origin of these flows should be different. In the heated substrate case, buoyancy drives the flow. In the isothermal case, vapor concentration gradient between the droplet interface and far field triggers the gas circulation. Apart from the simulations with test chamber, additional simulations with open boundaries are carried out to assess the boundary effect on evaporation. Results demonstrate that evaporation rate can change by up to 5.3\% in the absence of physical boundaries. Details of additional simulations are provided in Section~A of Supplementary Material.   

Transport mechanisms severely vary in the droplet and near gas region depending on the presence of thermocapillarity. Resultant velocity and temperature fields together with the superimposed energy flux and velocity streamlines are reported for substrate temperature of \mbox{65.4\unit{$^{\circ}$C}} with and without thermocapillarity in \cref{fgr:close_T}, whereas those for isothermal substrate are provided in Section~B of Supplementary Material for brevity. When thermocapillarity is accounted for, strong surface velocities shape the velocity field (\cref{fgr:close_T}a). At relatively higher contact angles ($\theta>44^{\circ}$), a single large convection cell is created by the surface flow from the contact line to the apex, the direction of decreasing interface temperature, in accordance with previous predictions \cite{hu2005}. The orientation of this vortex structure is designated as CCW, based on the right side positioned images, which include streamlines in the droplet domain, in \cref{fgr:close_T}a. Around $\theta=44^{\circ}$, the droplet experiences a transition period, during which tiny vortex structures arise near the contact line or apex or sometimes none at all during iterations, similar to \textit{the oscillatory transition convection period} reported in \cite{shi2017}. At smaller contact angles, B\'enard-Marangoni instability creates steady convection cells \cite{shi2017}. With decreasing contact angle, two behaviors are captured: i) directions of vortices may switch while the number of vortices remain the same and ii) number of vortices may increase. Generation of convection cells continues up to a critical contact angle, where the strong radial flow, known as the capillary flow generating coffee ring (stain) effect \cite{deegan1997}, dominates the instabilities. The onset is found between $\theta=1^{\circ}-2^{\circ}$ for the configuration demonstrated in \cref{fgr:close_T}. When thermocapillarity is not accounted for, the interplay between buoyant and radial flows shapes the velocity field. At relatively higher contact angles, a large single Rayleigh convection cell is created by buoyant forces in CW direction relying on the right side positioned images, which include streamlines in the droplet domain, in \cref{fgr:close_T}b. In 3-D, this convection cell creates an axisymmetric toroidal flow pattern, which was confirmed by previous experimental works \cite{dash2014buoyancy,he2016}. At relatively smaller contact angles, it is well known that the radial flow, which originates from the droplet surface and moves towards the contact line, becomes effective \cite{hu2005microfluid}. Transition between these two flow structures; however, has not been shown explicitly in previous studies. In the current work, this transition mechanism is clearly demonstrated. First, radial flow becomes apparent (around $\theta=48^{\circ}$). Then it grows, while the single Rayleigh convection cell shrinks ($28^{\circ}<\theta<48^{\circ}$). Finally, radial flow damps the Rayleigh convection completely at the contact angle of $24^{\circ}$. After that, radial flow is effective till the dryout.


\begin{figure}[h!]
\includegraphics[scale=1.2]{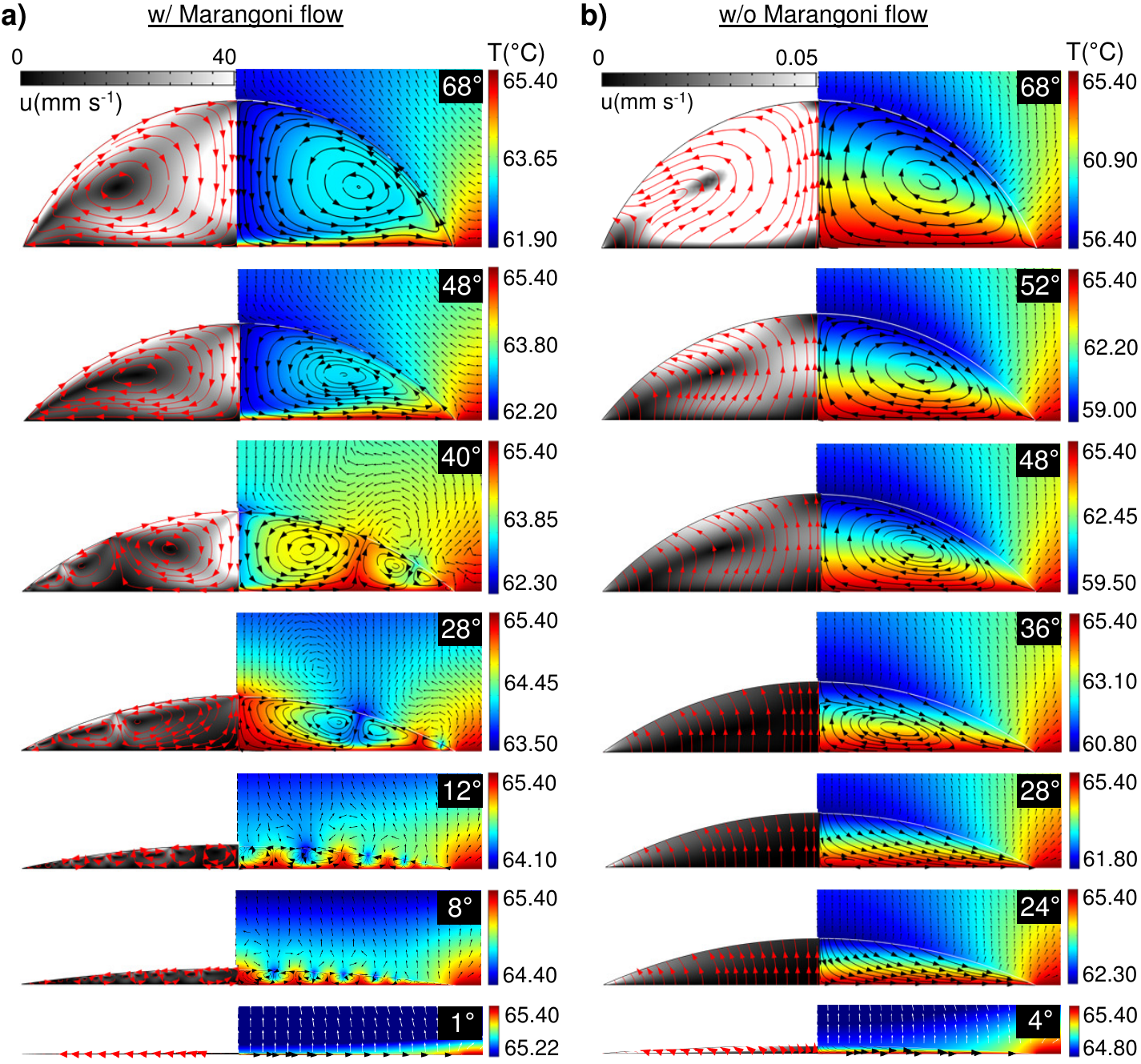}
\centering
\caption{Velocity magnitude field (left images) inside the droplet with superimposed total energy flux streamlines and temperature field (right images) inside the droplet with superimposed velocity streamlines and in the near droplet gas region  with superimposed normalized velocity vectors in the \textbf{a)} presence and \textbf{b)} absence of thermocapillarity. Substrate temperature is \mbox{65.4\unit{$^{\circ}$C}}. White lines indicate the liquid-vapor interface. The value of the corresponding contact angle is specified on each plot. Note that velocity magnitude scale bars in \textbf{a)} and \textbf{b)} are common for the corresponding plots, whereas individual temperature scale bars are utilized for each plot.}
\label{fgr:close_T}
\end{figure}

The resultant internal velocity fields are strongly coupled with the energy transport routes as shown in left sided images in \cref{fgr:close_T}. When Marangoni flow is present, velocities inside the droplet are nearly three orders of magnitude higher than those in the case omitting the thermocapillary effect. Consequently, energy transport paths follow the velocity streamlines by manifesting the convection as the primary energy transport mechanism. When Marangoni flow is absent, on the other hand, conduction accompanies convection to a greater extent because of the moderate internal velocities of buoyant convection. Velocity magnitudes further decrease with decreasing contact angle leading to increased conduction heat transfer. Upon the shrinking of the buoyant convection cell ($\theta<48^{\circ}$), conduction starts to dominate the convection in energy transport as demonstrated by non-stretched energy flux streamlines between the substrate and droplet surface.

\begin{figure}[h!]
\includegraphics[scale=0.8]{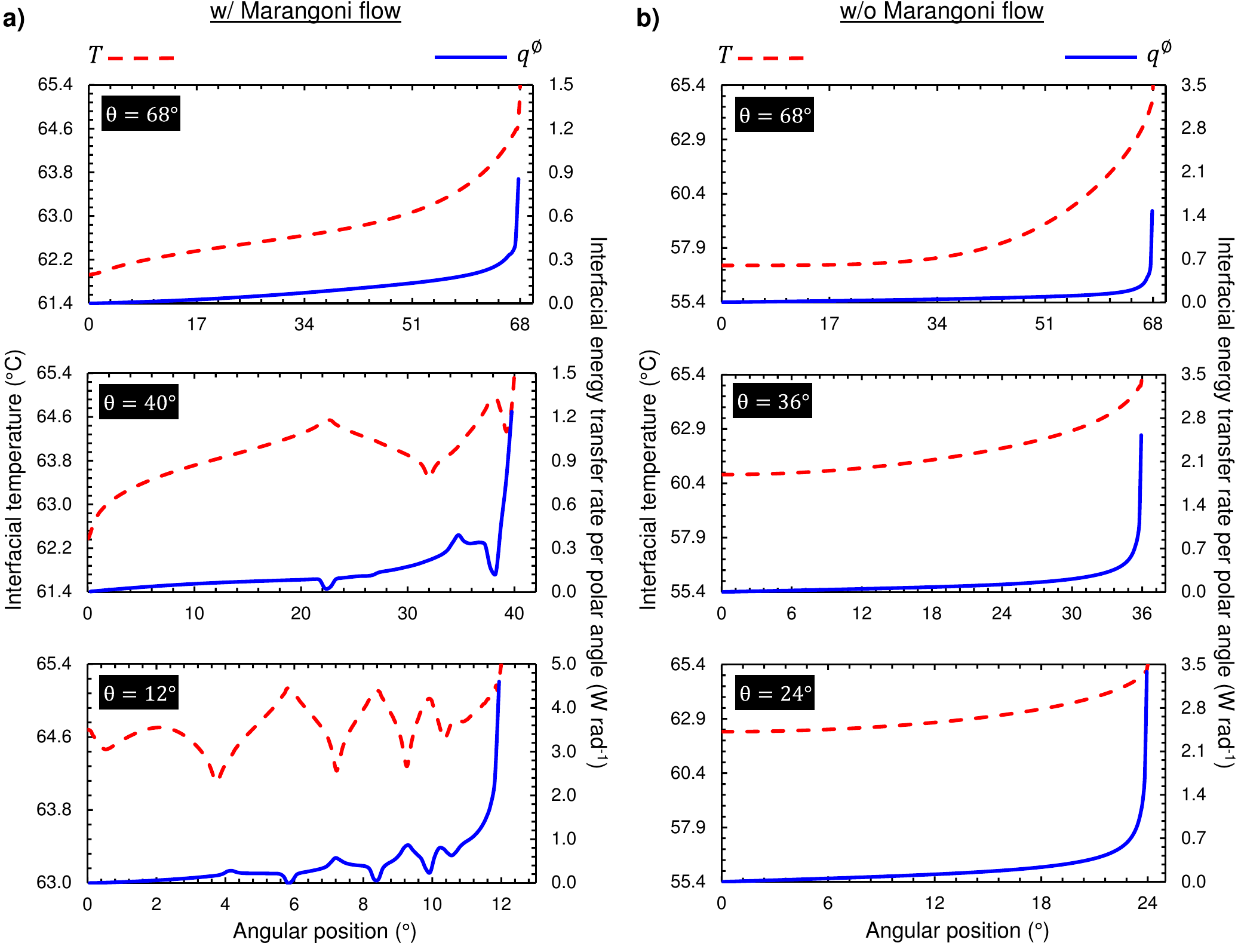}
\centering
\caption{Distributions of interfacial temperature and energy transfer rate per polar angle \textbf{a)} with and \textbf{b)} without thermocapillarity at selected contact angles. Substrate temperature is \mbox{65.4\unit{$^{\circ}$C}}. Angular position starts at the apex and terminates at the contact line.}
\label{fgr:q_dist}
\end{figure}

On a heated substrate, a droplet is expected to have a decreasing interfacial temperature from the contact line towards the apex. This temperature variation can be monotonic or thermocapillarity \cite{zhang2014temperature} and boyancy driven \cite{akkus2020theoretical} flows may result in non-monotonic variation depending on the configuration of the problem such as the substrate conductivity, contact angle and the superheat. In the problem of interest, interfacial temperature was monotonically varying for all contact angles in the absence of Marangoni flow as shown in Figs.~\ref{fgr:close_T}b~and~\ref{fgr:q_dist}b. Because the resultant flow fields do not possess multiple convection cells. Similarly, at relatively higher contact angles ($\theta>44^{\circ}$), the formation of a single convection cell, albeit in the reverse direction, results in a monotonic interfacial temperature variation when the Marangoni flow is present (Figs.~\ref{fgr:close_T}a~and~\ref{fgr:q_dist}a). With the formation of multiple convection cells due to B{\'e}nard-Marangoni instability ($\theta<44^{\circ}$), the variation of the interface temperature becomes non-monotonic. When two reverse circulating vortices meet at the interface, a local temperature peak forms if the vortices carry the liquid from the substrate to the interface. On the contrary, a local temperature dip forms when adjacent vortices carry the liquid away from the interface. A conspicuous result is that some of these temperature dips become cooler than the apex temperature at relatively smaller contact angles ($\theta<36^{\circ}$) such that the apex becomes not the coolest region of the droplet any more. Finally, when radial flow replaces B{\'e}nard-Marangoni convection cells ($\theta=1^{\circ}$), apex temperature becomes the minimum and interfacial temperature is monotonic again.

Although the distribution of evaporation flux is commonly reported in the literature, evaporation is not the sole phenomenon contributing to the interfacial energy transfer. Especially, in case of heated substrates, buoyant flow of gas warms the droplet surface. In fact, variation of interfacial energy transfer rate  is affected by the evaporation rate and the energy transport mechanisms in both phases. The distributions of energy transfer rate along the droplet interface together with the interfacial temperatures are plotted for selected contact angles in \cref{fgr:q_dist}. Contrary to common practice of presenting flux values, we report the distribution of total rates (per polar angle) in order to include the effect of increasing interfacial area with the increasing polar angle. As expected, energy transfer rate diverges near the contact line due to decreasing conduction resistance of thinning film. Inversely, increasing conduction resistance and decreasing interfacial area result in minimum energy transfer rates at the apex. The variation between these extremities is monotonic in the absence of multiple vortices. When B{\'e}nard-Marangoni convection cells are present, the distribution becomes highly non-monotonic by the presence of the local extrema that form at the intersection point of two reverse circulating vortices. Except the regions near the contact line and apex, local extrema of temperature and energy transfer rate distributions have an inverse relation, that is, a point of minimum energy transfer rate appears at the peak temperature points (or vice versa). At first glance, this result is perplexing since the elevated temperature could be expected to raise the local evaporation rate, the total energy transfer rate thereof. However, the evaporation is actually suppressed by the gas flow towards the interface. The gas flow forms as a result of the interaction of the velocity fields of two phases. More specifically, strong surface flows that split off at this point pull the contiguous gas molecules in the opposite directions. Consequently, the drawn gas molecules are replenished by the normal flow of the gas towards the interface, which suppresses the evaporation at this point. On the other hand, the inverse mechanism creates a point of maximum energy transfer rate at the point of minimum temperature. Overall, these mechanisms create adjacent vortices in the gas phase, but in the opposite direction of those in the liquid phase.  

Direct visualization of Marangoni flow in water is known to be challenging because of the sensitivity of water to surface contamination \cite{hu2005}. Alternatively, the presence of Marangoni flow can be assessed by examining the gas phase near the droplet surface, since liquid-gas interaction significantly alters the gas dynamics near the interface as demonstrated in \cref{fgr:close_T}a. Identification of the Marangoni flow \textit{via} gas phase may be difficult at higher contact angles because of the upward oriented gas flow. However, at smaller contact angles, gas vortices associated with the B{\'e}nard-Marangoni instability can be distinctively captured by a proper air visualization technique or any alternative methods such as the instant measurement of vapor concentration \cite{zhang2021}. Distributions of the near interface vapor concentration are provided in Fig.~S3 of Supplementary Material for the selected cases with B{\'e}nard-Marangoni instability.  

\begin{figure}[h!]
\includegraphics[scale=1.0]{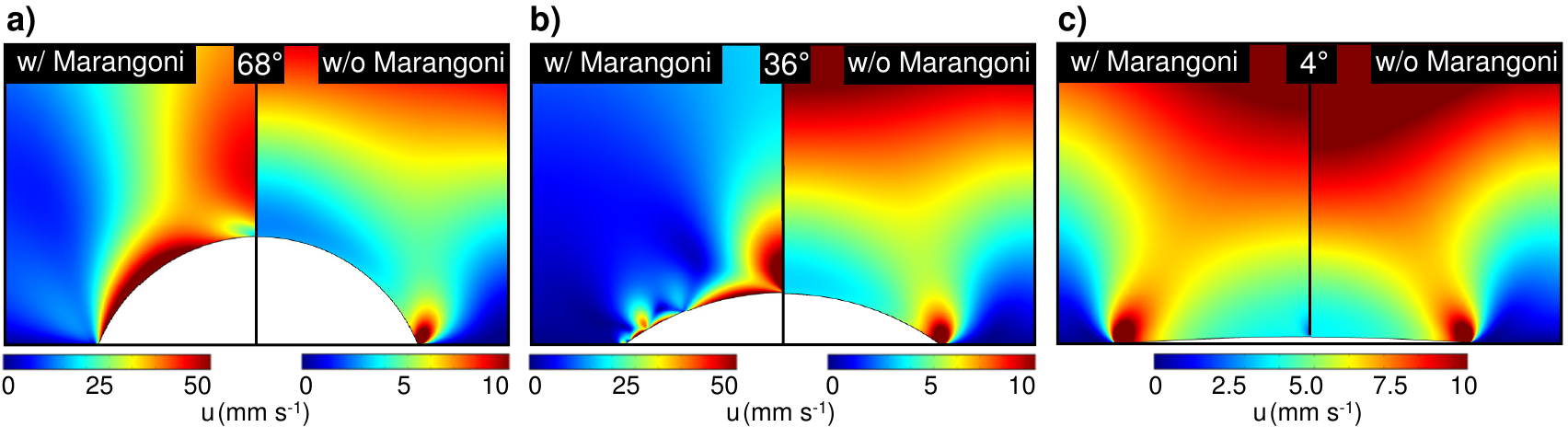}
\centering
\caption{Velocity magnitude field in the near droplet gas region in the presence and absence of thermocapillarity at the contact angles of \textbf{a)} 68$^{\circ}$, \textbf{b)} 36$^{\circ}$, and \textbf{c)} 4$^{\circ}$. Substrate temperature is \mbox{65.4\unit{$^{\circ}$C}}.}
\label{fgr:stefan}
\end{figure}

A remarkable feature of the current work is the inclusion of Stefan flow, which has been widely adopted in fuel droplet evaporation studies, but remained restricted in studies focused on the sessile water droplet evaporation. In the model presented, normal component of the interfacial gas velocity is determined based on the Stefan flow of air. Consequently, near surface gas flow field is shaped by the Stefan flow as shown in \cref{fgr:stefan}. In the absence of themocapillarity, normal component of the interfacial velocity dominates its tangential counterpart and the divergent evaporation flux near the contact line creates a distinctive gas jet. When thermocapillarity is accounted for, strong surface velocities outweigh the normal component and jet-like Stefan flow is not observable as seen in \cref{fgr:stefan} for the cases with contact angles $68^{\circ}$ and $36^{\circ}$. However, with the weakening Marangoni flow, jet-like Stefan flow becomes apparent (see $\theta=4^{\circ}$).      

\section{Conclusion}
\label{sec:conclusion}

Evaporation of a pinned sessile water droplet is modeled using temperature dependent thermophysical properties and incorporating all of the pertinent transport mechanisms in both phases. Predictions of the model excellently match with results of previous experiments. Interplay of transport mechanisms is highlighted in the presence and absence of thermocapillarity. When thermocapillarity is accounted for, surface tension driven flows (single Marangoni convection cell or multiple B{\'e}nard-Marangoni convection cells) are responsible for the mass and energy transport inside the droplet during most of the droplet lifetime. Then radial flow replaces thermocapillarity-induced flow. When thermocapillarity is omitted, at relatively higher contact angles, buoyancy-induced Rayleigh convection is responsible for the mass transport inside the droplet. Yet this convection is insufficient for the energy transport except the cases with high superheat values. With decreasing contact angle, radial flow starts to suppress buoyancy-induced flow by reducing the size of the Rayleigh convection cell. This transition period is depicted for the first time in the literature. Finally, radial flow completely damps the buoyancy-induced flow. We believe that identified transport mechanisms and their interaction show the potential to disclose physical mechanism of droplet evaporation in numerous applications from ink-jet printing to DNA stretching.
 
\addcontentsline{toc}{section}{References}

\bibliographystyle{unsrt}

\bibliography{references}

\section*{Declarations of interest}
\addcontentsline{toc}{section}{Competing_interests}
None.

\end{document}